\documentclass[11pt,a4paper]{article}

\usepackage{amsmath,amssymb}
\usepackage{epsfig,graphicx}
\usepackage{subfigure}
\usepackage{graphicx}
\usepackage{rotating}
\usepackage{cancel}
\usepackage{bm}
\usepackage{color}
\usepackage{comment}
\usepackage{cite}

\renewcommand\({\left(}
\renewcommand\){\right)}
\renewcommand\[{\left[}

\def\beq{\begin{equation}}
\def\eeq{\end{equation}}

\begin{document}
\numberwithin{equation}{section}
\title{{\normalsize  \mbox{}\hfill DESY 12-227; MPP-2012-158}\\
\vspace{2.5cm} 
\Large{\textbf{Searching for WISPy Cold Dark Matter with a Dish Antenna
\vspace{0.5cm}}}}

\author{Dieter Horns$^1$, Joerg Jaeckel$^{2,3}$, Axel Lindner$^4$, Andrei Lobanov$^{5,\star}$,\\ Javier Redondo$^{6,7}$, Andreas Ringwald$^4$\\[2ex]
\small{\em $^1$Institute for Experimental Physics, University of Hamburg, Luruper Chaussee 149, D-22761 Hamburg, Germany}\\[0.5ex]
\small{\em $^2$Institut f\"ur theoretische Physik, Universit\"at Heidelberg, Philosophenweg 16, 69120 Heidelberg, Germany}\\[0.5ex]
\small{\em $^3$Institute for Particle Physics Phenomenology, Durham DH1 3LE, United Kingdom}\\[0.5ex]
\small{\em $^4$Deutsches Elektronen Synchrotron DESY, Notkestrasse 85, D-22607 Hamburg, Germany}\\[0.5ex]
\small{\em $^5$Max-Planck-Institut f\"ur Radioastronomie, Bonn, Germany}\\[0.5ex]
\small{\em $^6$Arnold Sommerfeld Center, Ludwig-Maximilians-Universit\"at, Munich, Germany}\\[0.5ex]
\small{\em $^7$Max-Planck-Institut f\"ur Physik, Munich, Germany}\\[0.5ex]
\small{\em $^{\star}$Visiting Scientist, Universit\"at Hamburg/DESY}\\[0.5ex]
}

\date{}
\maketitle

\begin{abstract}
\noindent
The cold dark matter of the Universe may be comprised of very light and very weakly interacting particles, so-called WISPs.
Two prominent examples are hidden photons and axion-like particles. 
In this note we propose a new technique to sensitively search for this type of dark matter with dish antennas. 
The technique is broadband and allows to explore a whole range of masses in a single measurement.
\end{abstract}

\newpage

\section{Introduction}
One of the most important open questions in physics is the nature of dark matter. Among the well-motivated
candidates from particle physics are axions and other very weakly interacting slim particles (WISPs)~\cite{Jaeckel:2010ni,Ringwald:2012hr}, e.g. axion-like particles (ALPs) or hidden photons (HPs). Recently, it has been shown that, both for ALPs and HPs whose dominant interactions with the standard model constituents arise from couplings to photons, a huge region in the parameter spaces spanned by the coupling strength to photons and the ALP or HP mass can give rise to the observed cold dark  matter~\cite{Arias:2012az,Nelson:2011sf}. The parameter regions in which this is a viable option are shown as the red shaded areas in 
Figs.~\ref{fig:HP} and \ref{fig:ALP}.

Due to their small mass and their prominent interaction with 
photons, ALPs and HPs require entirely different detection methods than weakly interacting massive particles (WIMPs).
While WIMPs are searched for in scattering experiments, WISP hunts usually rely on the conversion of WISPs to photons.  
Building on recent and current experiments~\cite{Zioutas:1998cc,Moriyama:1998kd,Robilliard:2007bq,Jaeckel:2007ch,Zavattini:2007ee}, helioscopes~\cite{Sikivie:1983ip,Redondo:2008aa} like TSHIPS~\cite{Schwarz:2011gu} or the proposed IAXO~\cite{Irastorza:2011gs,Irastorza:2012qf} and light-shining-through-a-wall experiments~\cite{Okun:1982xi,Redondo:2010dp} like ALPS-II or REAPR can probe sizable parts of the relevant parameter space in the foreseeable future. 
Those experiments use WISPs produced in the centre of the Sun or in the laboratory, and therefore do not rely on WISPs being dark matter. 
Whilst this makes them independent of the assumption that WISPs form all (or a sizable part of) dark matter, they ignore a plausible copious source of WISPs and address only indirectly the issue of identifying the nature of the dark matter.

Haloscopes~\cite{Sikivie:1983ip} such as ADMX~\cite{Asztalos:2009yp,Asztalos:2011ei} (or CARRACK II~\cite{Tada:1999tu}) on the other hand are direct dark matter searches exploiting microwave cavities\footnote{Other proposals use optical cavities~\cite{Melissinos:2008vn}. Finally, an alternative technique relying on the QCD axion coupling to neutrons has recently been proposed in~\cite{Graham:2011qk}.}. They achieve enormous sensitivity by enhancing the conversion of WISPy dark  matter 
into photons in a resonant cavity. The drawback of this technique is that the cavity frequency has to be equal (within the bandwidth) to the energy of the dark matter particles which is essentially given by their mass. Since unfortunately the mass is not known 
a-priori\footnote{In some models the mass of an axion required to be 
most of the dark matter can be determined up to uncertainties in the instantonic QCD potential. 
However, if the dominant fraction of axion dark matter arises after inflation from the misalignment mechanism, the 
dark matter density is proportional to the (square of) the misalignment angle which is determined by chance and can 
easily vary by an order of magnitude without fine-tuning. This leads to uncertainties of a similar size in the axion mass.}, 
the operation of these experiments requires a relatively slow scan over frequencies. 
WISPy dark matter is possible in a wide range of masses~\cite{Arias:2012az}.
Therefore, it is clearly advantageous to have a broadband experiment to scan fast and easily as much parameter space as possible. 
 
In this paper, we propose such a broadband search, using the electromagnetic radiation emitted by conducting surfaces when excited by 
cold dark matter ALPs or HPs. The idea is to focus this radiation into a detector by using a conveniently shaped surface, a spherical cap. 
This way ``dish antennas'' coupled to a sensitive broadband receiver turn out to be excellent dark matter detectors. 
We find that dish antennas can compensate the lack of a resonant enhancement, used in haloscopes, by using a large surface area. 
For HPs where no magnetic field is required for the conversion
this is very easy to realize and large antennas even allow to exceed the sensitivity of dark matter searches with resonant cavities.
For the ALP case the size is limited by the need to embed the whole dish into a magnetic field.
Therefore, as we will later see, in this case dish antennas are most advantageous to explore higher masses where the same enhancement requires a 
smaller area and cavity experiments loose sensitivity because of the small volume of the cavity. 
In addition to this we stress that, given a suitable detector, dish antenna searches require no scanning and 
the technique can be used in a large range of masses from below $\mu$eV to beyond eV.

The paper is organized as follows: in Sec.~\ref{sec:rad} we characterize the radiation emitted by a conducting surface when excited by cold dark matter HPs. ALPs are considered in
Sec.~\ref{sec:ALP}. In Sec.~\ref{sec:boundaries} we discuss effects of the environment, like magnetic field boundaries, shielding and non-vacuum condition. We conclude in Sec.~\ref{sec:con}.

\section{Dish antenna search for hidden photons}\label{sec:rad}

Hidden photons ($\tilde{X}^{\mu}$ with field strength $\tilde{X}^{\mu\nu}$) interact with ordinary photons ($A^{\mu}$ with field strength $F^{\mu\nu}$) via a so-called kinetic mixing~\cite{Holdom:1985ag},
\begin{equation}
\label{HPlagrangian0}
\mathcal L= -\frac{1}4F_{\mu\nu}F^{\mu\nu}-\frac{1}4\tilde{X}_{\mu\nu}\tilde{X}^{\mu\nu}-\frac{\chi}2 F_{\mu\nu} \tilde{X}^{\mu\nu}+\frac{m_{\gamma'}^2}2 \tilde{X}_\mu \tilde{X}^\mu+ J^\mu A_\mu,
\end{equation}
where $\chi$ is the dimensionless parameter quantifying the kinetic mixing and $J^{\mu}$ is the ordinary electromagnetic current. Typical values predicted in fundamental extensions of the Standard Model for $\chi$ are in the range $10^{-12}-10^{-3}$~\cite{Dienes:1996zr}.

\begin{figure}[t]
   \centering
   \includegraphics[width=12cm]{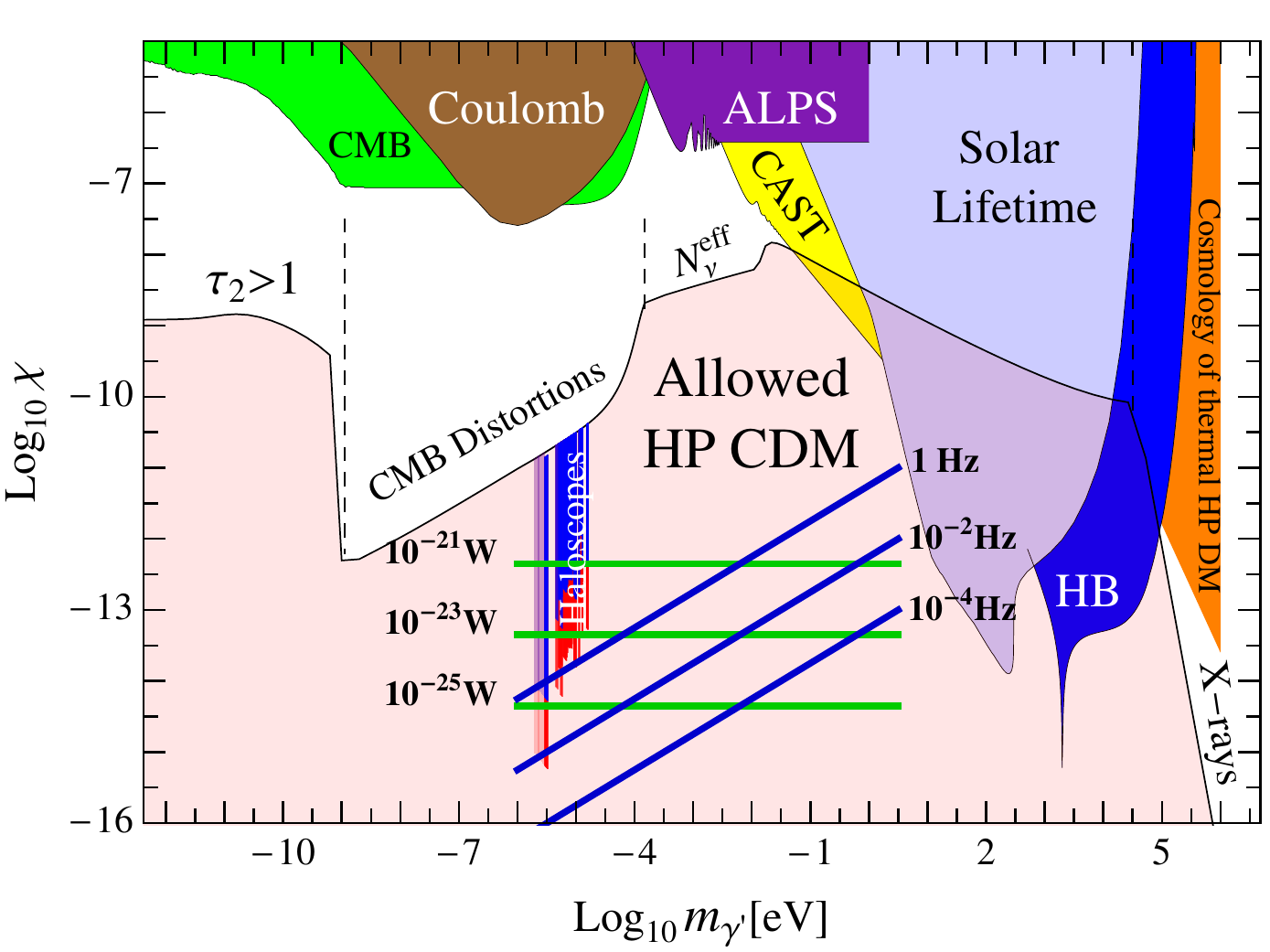} 
   \caption{The allowed parameter space for hidden photon cold dark matter (HP CDM) is shown in red (for details see Ref.~\cite{Arias:2012az}).  The regions in various colours are excluded by experiments and astrophysical observations that do not require HP dark matter (for reviews see~\cite{Ringwald:2012hr,Jaeckel:2010ni}). The lines correspond to the sensitivity of a dish antenna ($1\,{\rm m}^{2}$) search with a detector sensitive to $10^{-21}$, $10^{-23}$ and $10^{-25}$\,{\rm W} (green, from top to bottom) and $1$, $0.01$ and $10^{-4}$ photons per second (blue, from top to bottom).}
   \label{fig:HP}
\end{figure}

The parameter range of interest for hidden photon dark matter can be inferred from Fig.~\ref{fig:HP}. 
It is quite broad in mass and thus a broadband search for its signature seems appropriate. 

It is convenient to perform a shift $\tilde{X}^{\mu}\rightarrow X^{\mu}-\chi A^{\mu}$ which eliminates the kinetic mixing term.
After this shift and dropping terms of order $\sim \chi^2$, when convenient\footnote{We will continue to do so in the following.},  we have the Lagrangian,
\begin{equation}
\label{HPlagrangian}
\mathcal L= -\frac{1}4F_{\mu\nu}F^{\mu\nu}-\frac{1}4 X_{\mu\nu} X^{\mu\nu}+\frac{m_{\gamma'}^2}2 (X_\mu X^\mu-2\chi A_{\mu}X^{\mu}+\chi^2 A_{\mu}A^{\mu})+ J^\mu A_\mu.
\end{equation}
Let us consider the equation of motion for plane waves with frequency $\omega$ and momentum $k$, 
\begin{equation}
\left[(\omega^2-k^2)\left(
\begin{array}{cc}
1  & 0     \\
0 &  1   \\  
\end{array}
\right)
-
m^{2}_{\gamma^{\prime}}\left(
\begin{array}{cc}
\chi^2 & -\chi     \\
 -\chi &  1    \\  
\end{array}
\right)
\right]
\left(
\begin{array}{c}
\mathbf{A}\\
\mathbf{X}
\end{array}
\right)=
\left(\begin{array}{c}
0\\
0
\end{array}
\right),
\end{equation}
where we have used that $X^{0}=A^{0}=0$ can be achieved with a suitable gauge choice.
In this basis the dark matter solution corresponds to the spatially constant mode\footnote{In our galaxy dark matter particles are expected to have a small velocity $\sim10^{-3}$. Accordingly we have a momentum $|\mathbf{k}_{0}|\sim 10^{-3}m_{\gamma^{\prime}}$. However, for our consideration this is negligible in first approximation.
Going beyond this simple approximation, one finds that the momentum of the dark matter HPs parallel to the plane adds to the momentum of the outgoing photon. This leads to small angular corrections to ordinary photon emission from the surface. Given a detector with a sufficient spatial resolution this could be used to determine directional information on the velocity distribution of dark matter.}
 $k=0$, oscillating with frequency 
$\omega=m_{\gamma^{\prime}}$,
\begin{equation}
\left(
\begin{array}{c}
\mathbf{A}\\
\mathbf{X}
\end{array}
\right)
\bigg|_{\rm{DM}}=
\mathbf{X}_{\rm{DM}}
\left(
\begin{array}{c}
-\chi\\
1
\end{array}\right)\exp(-i\omega t).
\end{equation}

In the following we will consider two possibilities for the direction of the vector field $\mathbf{X}_{\rm DM}$. 
\begin{itemize}
\item[(i)]{$\mathbf{X}_{\rm{DM}}$ is the same everywhere in space.}
\item[(ii)]{The hidden photons behave like a gas of particles with random directions, i.e. we have a mixture of hidden photons ``pointing'' in random directions.}
\end{itemize}

If we suppose that all the dark matter in the galactic 
halo, whose energy density is of order 
\begin{equation}
\rho_{\rm CDM, halo}\sim \frac{0.3\,{\rm GeV}}{\rm cm^3}, 
\end{equation}
is comprised of a hidden photon condensate, then the energy density in hidden photons is given by
\begin{equation}
\label{dmcond}
\rho_{\rm HP}=\frac{m^{2}_{\gamma^{\prime}}}{2}\langle|{\mathbf{X}}_{\rm{DM}}|^2\rangle=\rho_{\rm CDM, halo}\sim\frac{0.3\,{\rm GeV}}{\rm cm^3}.
\end{equation}
The average is only relevant in the case (ii) where we treat the hidden photons as a gas. In case (i) it is trivial.

A small fraction of the energy density is, however, also present in the form of a 
small oscillating ordinary electric field (where the average is taken over a sufficiently large spatial volume),
\begin{equation}
\mathbf{E_{\rm DM}}=-\partial_{0}\mathbf{A}=\chi m_{\gamma^{\prime}} \mathbf{X}_{\rm{DM}}.
\end{equation}
Using Eq.~\eqref{dmcond} we find that this corresponds to an (oscillating) electric field of amplitude
\begin{equation}
\sqrt{\langle |\mathbf{E_{\rm DM}}|^2\rangle}=\chi \sqrt{2\rho_{\rm{CDM, halo}}} \sim 3.3\times 10^{-9}\ \frac{V}{\rm m} \left( \frac{\chi}{10^{-12}}\right)\left(\frac{\rho_{\rm{CDM, halo}}}{0.3\,{\rm GeV}/{\rm{cm}}^3}\right),
\end{equation}
which oscillates with a frequency
\begin{equation}
f =\frac{m_{\gamma^\prime}}{2\pi} = 0.24\ {\rm GHz}\ \left( \frac{m_{\gamma^\prime}}{\rm \mu eV}\right) .
\end{equation}

To see how a dish antenna can be used for a WISP search let us first consider the effects of an infinite plane mirror at $x=0$.
In particular let us look at the fields on the right hand side of the plane.
Far away from any boundaries the HP dark matter has the following visible and hidden electric fields,
\begin{equation}
\left(\begin{array}{c}
\mathbf{E}(\mathbf{x},t)\\
\mathbf{E}_{\rm hid}(\mathbf{x},t)
\end{array}\right)\bigg|_{\rm DM}
=- m_{\gamma^{\prime}} \mathbf{X}_{\rm DM}
\left(
\begin{array}{c}
-\chi    \\ 
1  
\end{array}
\right)\exp(-i\omega t),
\end{equation}
where
\begin{equation}
\omega=m_{\gamma^{\prime}}.
\end{equation}

If we consider the plane to be a perfect mirror for the frequency of interest the boundary condition for the ordinary electric field is
\begin{equation}
\mathbf{E}_{||}(x=0,y,t)=0.
\end{equation}
Here, the index $||$ denotes the directions parallel to the plane.

Ordinary matter (like the mirror) can only affect the ordinary electric field. 
The boundary condition on the plane is fulfilled on the right hand side of the plane by adding an outgoing plane wave with the same frequency,
\begin{equation}
\left(\begin{array}{c}
\mathbf{E}\\
\mathbf{E}_{\rm hid}
\end{array}\right)_{\rm out}
=\mathbf{E}_{\rm DM,||}\exp(-i(\omega t-\mathbf{x}\mathbf{k}))
\left(
\begin{array}{c}
1    \\ 
\chi
\end{array}
\right)
\end{equation}
where 
\begin{equation}
\mathbf{k}=\omega\left(1,0,0\right)^{\rm T}.
\end{equation}
In other words we have an outgoing (almost) ordinary electromagnetic wave emitted perpendicular to the surface. This is of course to be expected from a microscopical point of view: the electrons of the mirror's surface oscillate under the tiny electric field of the HP emitting a reflected electromagnetic wave of the same amplitude and opposite phase. 
Its polarization lies within the surface plane and the magnitude is set by the dark matter density via,
\begin{equation}
\sqrt{\langle |\mathbf{E}_{\rm DM, ||}|^{2}\rangle}=\chi \sqrt{2\rho_{\rm DM}}\,\alpha\quad\quad{\rm where}\quad\quad
\alpha=\bigg\{
\begin{array}{cl}
|\cos(\theta)| & {\rm case\,\, (i)}\\
\sqrt{\frac{2}{3}}& {\rm case\,\, (ii)}
\end{array}.
\end{equation}
Here $\theta$ is the angle between the HP field (for case (i) it points in the same direction everywhere) and the plane. In case (ii) the $\sqrt{2/3}$ arises from the average over the random orientation of the HPs.

On the $x=0$ plane, the dark matter field together with the outgoing wave then fulfill 
the boundary condition for the electric field components parallel to the plane,
\begin{eqnarray}
\label{nearplane}
\left(\begin{array}{c}
\mathbf{E}\\
\mathbf{E}_{\rm hid}
\end{array}\right)_{\rm total, ||}
&\!\!=\!\!&
\mathbf{E}_{\rm DM, ||}\,\left[
\left(\begin{array}{c}
1 \\
\chi
\end{array}\right)\exp(-i(\omega t-\mathbf{k}\mathbf{x}))+
\frac{1}{\chi}\left(\begin{array}{c}
-\chi\\
1
\end{array}
\right)\exp(-i\omega t)
\right]_{x=0}
\\\nonumber
&\!\!=\!\!&
\mathbf{\mathbf{E}_{\rm DM, ||}}\frac{1}{\chi}
\left(\begin{array}{c}
0\\
1
\end{array}
\right)
.
\end{eqnarray}
Evaluating the electric field at a finite distance from the plane we find
\begin{equation}
\label{electricnear}
\mathbf{E}(\mathbf{x},t)_{||}=\mathbf{E}_{\rm DM,||}\exp(-i\omega t)\left(\exp(i\mathbf{k}\mathbf{x})-1\right),
\end{equation}
which obviously fulfills $\mathbf{E}_{||}(\mathbf{x}=0,t)=0$.

For our detection scheme the crucial bit is that we have an outgoing (mostly) ordinary electromagnetic wave with electric field amplitude
$\mathbf{E}_{\rm DM,||}=\chi m_{\gamma^{\prime}}\mathbf{X}_{\rm DM}$ and direction perpendicular to the plane. 
This suggests using a spherical mirror which focuses the waves in its centre. 
The simplest and most practical case is when this mirror covers only a small part of the sphere and this is the case we will look at in the following. 
For practical purposes this is nothing but a dish antenna\footnote{Most dish antennas are not spherical, but parabolic. However, since they are usually very flat they can be taken as approximately spherical. 
Violations of sphericity can be compensated by increasing the size of the detector or by refocusing the signal.}.

\begin{figure}[t] 
   \centering
   \includegraphics[width=2in]{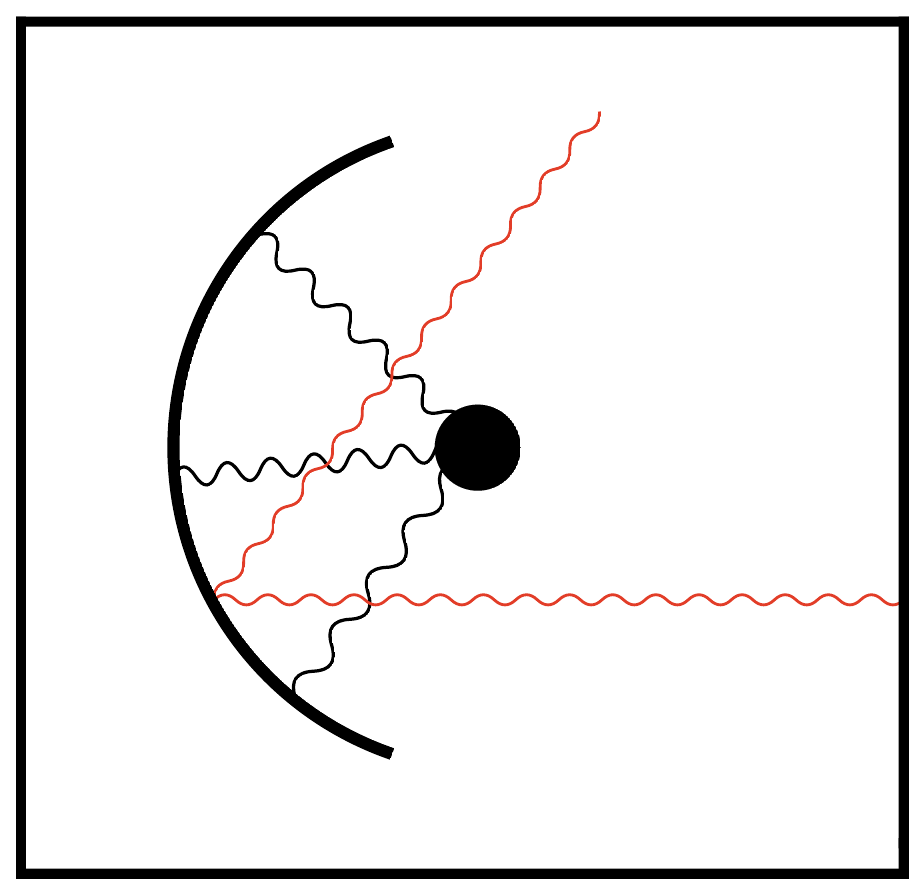} 
   \caption{Sketch of our WISPy cold dark matter experiment. Non relativistic HPs or ALPs mixing with photons are converted into monochromatic photons (black) emitted from the surface of an spherical dish antenna and focused in the centre, where a broadband detector is placed. Photons emitted from other boundaries or from far away sources (red) are typically not focused there.}
   \label{fig:scheme}
\end{figure}

Let us now consider the situation where the wavelength of the outgoing wave is much smaller than the radius of the dish antenna,\begin{equation}
r_{\rm dish}\gg \lambda=\frac{2\pi}{\omega}=\frac{2\pi}{m_{\gamma^{\prime}}}.
\end{equation}
In this case diffraction effects are negligible and we can use ray-optics as a good approximation. 
In this approximation we can easily see the effect an arbitrary shaped mirror: \emph{all rays are emitted perpendicular to the 
surface}. It is then obvious that the best focussing occurs in the \emph{centre} of a spherical mirror. 
Our experimental proposal, outlined in Fig.~\ref{fig:scheme}, consists therefore in using an spherical dish antenna with a broadband detector placed in its centre. 

The power concentrated in the centre of such a mirror 
is then approximately 
\begin{equation}
\label{concentrated}
P_{\rm center}\approx A_{\rm dish}  \langle |\mathbf{E}_{\rm DM,||}|^2\rangle=\langle\alpha^2\rangle_{\rm dish} \chi^2\rho_{\rm CDM} A_{\rm dish},
\end{equation}
where $A_{\rm dish}$ is the surface area of the mirror and the average is taken over the surface of the dish.  
In this expression we have neglected the small electric field due to the dark matter HP field in the center (that exists regardless of the mirror) 
because it is not focused and therefore smaller.

Let us compare the power $P_{\rm centre}$ to the power output of a resonant cavity experiment~\cite{Arias:2012az},
\begin{equation}
\label{cavity}
P_{\rm resonant \,\,cavity}=\kappa \chi^2 m_{\gamma^{\prime}}\rho_{\rm CDM} Q V_{\rm cavity} {\mathcal{G}}_{0}(\alpha^{\prime})^2.
\end{equation}
Here $Q$ is the quality factor of the cavity, $V_{\rm cavity}$ its volume and ${\mathcal{G}}_{0}$ is a geometrical factor which encodes the overlap between the cavity mode and the dark matter field. Finally, $\alpha^{\prime}$ is $|\cos(\theta)|$ for model (i) and $\sqrt{1/3}$ for case (ii)~\cite{Arias:2012az}. Finally, $\kappa$ is the coupling of the cavity mode to the 
detector which, for good coupling, should not be too far away from 1.

Comparing Eqs.~\eqref{concentrated} with \eqref{cavity} we find that up to some geometrical factors $QV_{\rm cavity}$ is replaced by $A_{\rm dish}/m_{\gamma^{\prime}}\sim A_{\rm dish}\lambda$. 
Having lost the $\sim Q\sim 10^6$ enhancement due to the resonance we can attempt to compensate for this by using a large area of the dish antenna.
In particular for not too large wavelengths this becomes a very good option considering that the
typical volume of a cavity is of the order of $\sim\lambda^3\sim 1/m^{3}_{\gamma^{\prime}}$. Using this we find that the lost quality factor can be (more than) compensated\footnote{Note that it is in principle possible to construct cavities with not too small geometrical factors that have large volumes at a fixed frequency~\cite{Baker:2011na}. One possibility is to have one or two sides large compared to the wavelength, which also has other advantages, see~\cite{Irastorza:2012jq}. However, going to extreme ratios is technically quite challenging.} by a large $A_{\rm dish}/\lambda^2$
which plays the role of an ``effective'' $Q$,
\begin{equation}
Q_{\rm eff}\sim A_{\rm dish}/\lambda^2 \sim 10^{4}\left(\frac{A_{\rm dish}}{1\,{\rm m^{2}}}\right)\left(\frac{\lambda}{{\rm cm}}\right)^{2}. 
\end{equation}

How do we get the enhancement of the output power? Roughly speaking we have an effective volume of $\sim A_{\rm dish} \lambda $ because we get conversion on a surface of size $A_{\rm dish}$ and after a distance $\sim\lambda$ the hidden photon is basically converted.

We stress that the power is concentrated in the \emph{centre of the spherical dish} and \emph{not in the usual focus point} which, for a spherical dish antenna (which covers only a small fraction of a full sphere), lies halfway between the centre and the surface, cf.~Fig.~\ref{fig:scheme}.
Therefore, far-away background sources, which will emit radiation that arrives at the dish as approximate plane waves, will be concentrated in the focal point and do not get focused into the center detector. Of course, to reduce avoidable background it is surely best to put the whole setup, dish and detector in the centre in a shielding box\footnote{Note that the shielding itself can get excited by the HP dark matter field and emit electromagnetic radiation which can interfere with our measurements. We will briefly consider these effects in Sec.~\ref{sec:boundaries}, but from what we learned so far (this radiation will also be perpendicular to the box surface) we can already conclude that we can arrange the geometry in such a way that interference is minimized.}.
It should be stressed that although the experiment performs a broadband search the signal from HP dark matter would be a very narrow peak which should be easily distinguishable from background sources.

For a first experiment one could take a standard dish antenna and simply move the receiver/detector from the focal point to twice the distance. 

Finally an important (perhaps the most important) feature of this technique is that it is broadband. Given a detector with a suitably low background and high enough sensitivity, we can do a search for hidden photon over the whole frequency/mass range to which the detector is sensitive without the need to adjust the experiment. This is in stark contrast to a cavity experiment which achieves its $Q$ factor enhanced sensitivity only in a tiny range $\sim \omega/Q$ around the resonance frequency and for which one has to slowly scan through the desired mass range.  

Let us now estimate the sensitivity of such an experiment.
In principle the concentration mechanism is effective as long as diffraction is small, i.e. as long as the wavelength is much smaller than the size of the dish antenna (this is also the limit when the ray-approximation is reasonable). 
This limits us to masses
\begin{equation}
m_{\gamma^{\prime}}\gtrsim {\rm few}\times1\,\mu{\rm eV}\left(\frac{\rm m}{r_{\rm dish}}\right). 
\end{equation}
Aside from this the only limitations are:
\begin{itemize}
\item The dish should indeed provide a boundary condition of a vanishing electric field, i.e. it has to be a good reflector. For radio frequencies this can be achieved by using metal dishes. In the IR to near UV mirrors are obvious examples of reflective surfaces. Therefore we can easily cover a wide range of masses with this technique. 
\item The surface of the dish has to be smooth and well focused to the centre at length scales of the wavelength we want to probe, i.e. it can be taken as spherical to a good approximation. 
\item Thermal emission from the mirror provides a background for our measurement. In contrast to our signal it has a broad spectrum. Thermal emission is highly suppressed by the high reflectivity of the mirror and moreover it can be reduced by cooling the dish. Moreover, thermal radiation is emitted isotropically and will not be focused on the  detector. This translates into a relative suppression with respect to the signal of the order of the detector area divided by the dish area. 
\end{itemize}

In the radio frequency regime very small powers can be detected. Powers as low as $10^{-26}\,{\rm W}$ seem feasible and $10^{-23}\,{\rm W}$ are certainly possible.
Using Eq.~\eqref{concentrated} this can easily be translated into a sensitivity to the kinetic mixing parameter,
\begin{equation}
\label{hpsens1}
\chi_{\rm sens}=4.5\times 10^{-14}\,\left(\frac{P_{\rm{det}}}{10^{-23}\,{\rm W}}\right)^{\frac{1}{2}}\left(\frac{0.3\,{\rm GeV/cm^{3}}}{\rho_{\rm CDM, halo}}\right)^{\frac{1}{2}}
\left(\frac{1\,{\rm m^2}}{A_{\rm dish}}\right)^{\frac{1}{2}}\left(\frac{\sqrt{2/3}}{\alpha}\right).
\end{equation}
The corresponding limits for a $1\,{\rm m}^{2}$ dish are shown as the green lines in Fig.~\ref{fig:HP} for detectable powers of $P_{\rm det}=10^{-21},\,10^{-23}$ and $10^{-25}$~W (from top to bottom).

We note that $P_{\rm det}$ is the detectable power in presence of
various backgrounds such as the thermal background (see the third
bullet point above), the detector noise and unshielded environmental
backgrounds  ({\em e.g.}, sky brightness and ground
  noise). Reducing all those backgournds to the required level is
technically challenging but at least the two higher power levels are
certainly feasible in the radio frequency regime.  In particular,
  measurements in the 1--20 GHz range of frequencies (1.5--30 cm
  wavelengths) would be least affected by the sky brightness
  (dominated at these frequencies by the cosmic microwave background,
  with $T_\mathrm{sky} \approx T_\mathrm{CMB}$). At $\lambda>30$ cm,
  the sky brightness increases $\propto
  \lambda^{2.8}$~\cite{Kogut:2012}, while at $\lambda <1.5$\,cm, the
  atmospheric H$_2$O and O$_2$ lines will result in $T_\mathrm{sky} >
  100$\,K~\cite{Morrison:1977}. In both these regimes, shielding will
  become important for providing $P_\mathrm{det} < 10^{-23}\,{\rm W}$.

At higher frequencies it is often more sensible to think in terms of a detectable photon rate, $R_{\gamma,{\rm det}}$, instead of a detectable power,
\begin{equation}
\label{hpsens2}
\chi_{\rm sens}=5.6\times 10^{-12}\,\left(\frac{R_{\gamma, \rm{det}}}{1\,{\rm photon}/s}\right)^{\frac{1}{2}}
\left(\frac{m_{\gamma^{\prime}}}{\rm eV}\right)^{\frac{1}{2}}
\left(\frac{0.3\,{\rm GeV/cm^{3}}}{\rho_{\rm CDM, halo}}\right)^{\frac{1}{2}}
\left(\frac{1\,{\rm m^2}}{A_{\rm dish}}\right)^{\frac{1}{2}}\left(\frac{\sqrt{2/3}}{\alpha}\right).
\end{equation}
For a rate of one photon per $1$, $100$, and $10^{4}$~s the corresponding sensitivities are shown as the blue lines (from top to bottom) in Fig.~\ref{fig:HP}.
Of course, the blue and the green lines intersect when the powers/photon counting rates are equal.

At higher frequencies, in particular in the optical regime, thermal
backgrounds become negligible and shielding is easier, here the main
challenge is to have a sufficiently low dark count rate of the
detector.

We can clearly see from the plot that this technique could allow to explore huge areas in the hidden photon parameter space.

\section{Dish antenna search for axion-like particles}\label{sec:ALP}
\begin{figure}[t]
   \centering
   \includegraphics[width=12cm]{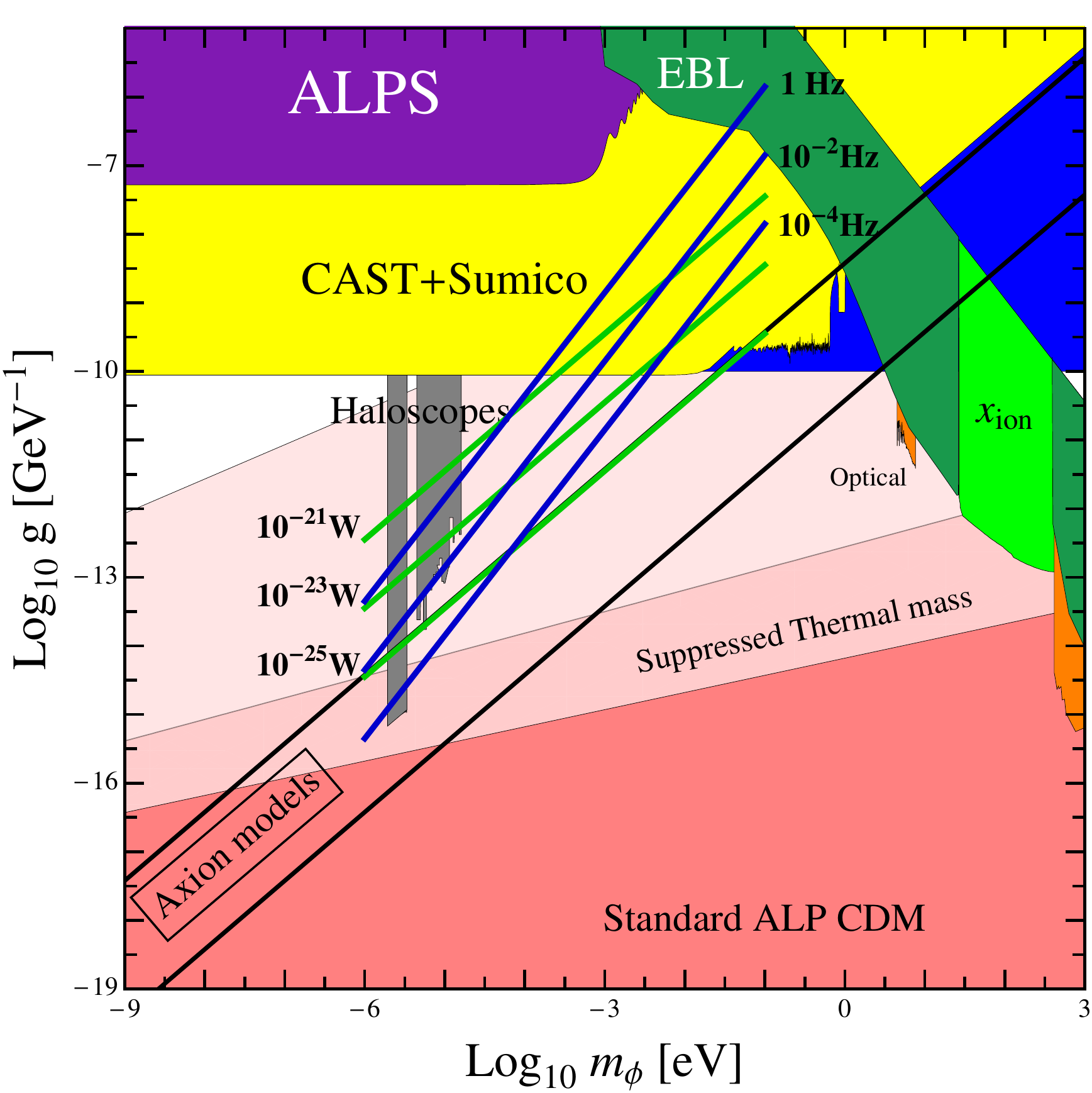} 
   \caption{The allowed parameter space for axion-like particle dark matter (ALP CDM) is shown in various shades of red (for details see Ref.~\cite{Arias:2012az}).  The various colored regions are excluded by experiments and astrophysical observations that do not require HP dark matter (for reviews see~\cite{Ringwald:2012hr,Jaeckel:2010ni}). The lines correspond to the sensitivity of a dish antenna ($1\,{\rm m}^{2}$ dish in a 5~T magnetic field) with a detector sensitive to $10^{-21}$, $10^{-23}$ and $10^{-25}$\,{\rm W} (green, from top to bottom) and $1$, $0.01$ and $10^{-4}$ photons per second (blue, from top to bottom).}
   \label{fig:ALP}
\end{figure}

Axion-like particles (ALPs) are pseudo-scalar particles coupled to two photons,
\begin{equation}
{\mathcal{L}}=-\frac{1}{4}F_{\mu\nu}F^{\mu\nu}+\frac{1}{2}\partial_{\mu}\phi\partial^{\mu}\phi-\frac{1}{2}m^{2}_{\phi}\phi^2 -\frac{g_{\phi\gamma\gamma}}{4}\phi F_{\mu\nu}\tilde{F}^{\mu\nu},
\end{equation}
where $\tilde{F}$ is the dual field strength.
The parameter space in the mass-coupling plane is shown in Fig.~\ref{fig:ALP} (for predictions from string theory see~\cite{axiverse}). 
The region where ALPs can be dark matter is shaded in red (the different densities correspond to different models described in~\cite{Arias:2012az}
the details of which are, however, not important for this note).

In presence of a magnetic background field $\mathbf{B}$ we have the linearized (in the photon field) equations of motion,
\begin{equation}
\label{alpmotion}
\left[(\omega^{2}-k^{2})
\left(
\begin{array}{cc}
1 & 0\\
0 & 1
\end{array}\right)
+\left(
\begin{array}{cc}
0 & -g_{\phi\gamma\gamma}|\mathbf{B}|\omega\\
-g_{\phi\gamma\gamma}|\mathbf{B}|\omega & m^{2}_{\phi}
\end{array}
\right)
\right]
\left(
\begin{array}{c}
\mathbf{A}_{||}\\
\phi
\end{array}
\right)
=
\left(
\begin{array}{c}
0\\
0
\end{array}
\right).
\end{equation}
Here $\mathbf{A}_{||}$ denotes the components of the photon field which are parallel to the magnetic field. The components $\mathbf{A}_{\perp}$ perpendicular to the magnetic field
are unaffected by the ALPs.

Let us consider the same situation as before with a reflecting plane. In addition we now have a magnetic field oriented parallel to the plane.
One can now easily check that the solutions to Eq.~\eqref{alpmotion} have the same form as those for the hidden photon case with the replacement,
\begin{equation}
\chi\rightarrow 
\frac{g_{\phi\gamma\gamma}|\mathbf{B}|}{m_{\phi}}.
\end{equation}
In particular after implementing the boundary condition of a vanishing electric field on the reflecting surface \emph{we have the same structure as in Eq.~\eqref{nearplane}}.
In analogy to Eqs.~\eqref{hpsens1} and \eqref{hpsens2} we therefore find,
\begin{equation}
\label{hpsens3}
g_{\phi\gamma\gamma,\,\,{\rm sens}}=\frac{3.6\times 10^{-8}}{\rm GeV}\left(\frac{5\,{\rm T}}{\sqrt{\langle |\mathbf{B}_{||}|^{2}\rangle}}\right)\left(\frac{P_{\rm{det}}}{10^{-23}\,{\rm W}}\right)^{\frac{1}{2}}
\left(\frac{m_{\phi}}{\rm eV}\right)\left(\frac{0.3\,{\rm GeV/cm^{3}}}{\rho_{\rm DM, halo}}\right)^{\frac{1}{2}}
\left(\frac{1\,{\rm m^2}}{A_{\rm dish}}\right)^{\frac{1}{2}},
\end{equation}
and
\begin{equation}
\label{hpsens4}
g_{\phi\gamma\gamma,\,\,{\rm sens}}=\frac{4.6\times 10^{-6}}{\rm GeV}\,\left(\frac{5\,{\rm T}}{\sqrt{\langle |\mathbf{B}_{||}|^{2}\rangle}}\right)\left(\frac{R_{\gamma, \rm{det}}}{1\,{\rm Hz}}\right)^{\frac{1}{2}}
\left(\frac{m_{\phi}}{\rm eV}\right)^{\frac{3}{2}}
\left(\frac{0.3\,{\rm GeV/cm^{3}}}{\rho_{\rm DM, halo}}\right)^{\frac{1}{2}}
\left(\frac{1\,{\rm m^2}}{A_{\rm dish}}\right)^{\frac{1}{2}},
\end{equation}
where $\langle |\mathbf{B}_{||}|^{2}\rangle$ is the average value of the magnetic field squared parallel to the antenna dish.

The sensitivities  for a $1\,{\rm m}^{2}$ dish are shown in Fig.~\ref{fig:ALP}. Green lines are again for the same values of fixed power and the blue lines are at fixed detectable photon rate.
The region that can be probed with such a setup is not quite as spectacular as in the hidden photon case. We note, however, that already with modest means a nevertheless quite sizeable area of new parameter space can be probed. In particular we can test masses $m_{\phi}\gtrsim {\rm few}\times 10^{-5}\,{\rm eV}$ which are more difficult to probe in experiments with resonant cavities. Moreover, we repeat that this can be done without a time consuming scan through the mass range.

\section{Non-ideal mirror and boundary effects}\label{sec:boundaries}
In this section we want to highlight a few points regarding the quality of the dish antenna and the magnetic field required for our proposed experiments. As we will argue below, they are typically not very relevant for the experiments we have exposed here, but can play a role at very low frequencies or in setups with different geometries. 

\subsection{Boundary effects}
Let us first consider the effects of the configuration of the magnetic field in the case of our experiment looking for axion-like particles. 
We have assumed so far that the extent of the magnetic field covers all the
relevant volume of the antenna and receiver. On the other hand, since the
electromagnetic radiation comes from the surface of the antenna we might think
that it is sufficient to cover only a small sliver of thickness $\sim \lambda$ around it.  
In order to assess how thick this slice has to be, let us
consider the flat antenna case once more. However, this time in
a  finite magnetic field region (extending from the mirror at $x=0$ to $x=L_B$). We can solve the
mixed equations of motion from $x=0$ to $x=L_B$ by using the boundary condition
${\mathbf E}_{||}(x=0)=0$ and match them with the solution outside of the
magnetic field region, which is trivial because there is no mixing.
Alternatively, let us compute the solution in a different way, which might
provide the reader with additional physical insight. Inspired by the perturbative
approach used in~\cite{Raffelt:1987im,Redondo:2010dp}, we take the mixing between the photon
and the ALP in a magnetic field as an interaction that converts one into the other. The amplitude per unit length for conversion is $i g |{\mathbf B}_{||}|/2$. 
The conversion can happen into photons that move to the left
or to the right, but the latter will bounce back in the reflecting surface and
interfere with the former, so we should consider the interference. Of course,
the two waves travel different distances so they have a phase difference.
Using a homogeneous ALP source $\phi=\phi_0 \exp({-i m_\phi t})$ we find that at a
distance $x>L_B$ from the mirror the electric field is 
\begin{eqnarray}
\nonumber
\mathbf{E}(\mathbf{x},t)_{||} &=& m_{\phi} \int_0^L dx' \, \phi \frac{i  g  {\mathbf B}_{||}}{2} 
\(\exp({i m_\phi (D-x')})-\exp({i m_\phi (D+x')})\)							\\[0.3cm]
&=&
\label{eq:outsideB}
\phi_0 \,g \, {\mathbf B}_{||}  \exp({-i m_\phi t})\exp({i m_\phi x}) 
(1-\cos(m_\phi L_B)) 
\end{eqnarray}  
(the minus sign in the second wave comes from the sign flip after the reflection in the mirror). 
If we compare it with the ALPs version of Eq.~\eqref{electricnear}, 
\begin{equation}
\label{electricnear2}
\mathbf{E}(\mathbf{x},t)_{||}=\frac{\sqrt{2\rho_{\rm CDM}}}{m_{\phi}}g\mathbf{B_{||}}\exp(-i\omega t)\left(\exp(i\mathbf{k}\mathbf{x})-1\right),
\end{equation}
with $\omega=m_\phi, \mathbf{k}=\omega(1,0,0)^{\rm T}$, 
we see that the factor $(e^{i {\mathbf k \mathbf x}}-1)$ has lost its constant
$-1$. In our calculation of Sec.~\ref{sec:rad} we saw that this originates from the little electric field that the dark matter ALP has \emph{in the background magnetic field}. 
Therefore, in the region of no magnetic field its absence is expected.
Additionally, a new factor $(1-\cos(m_\phi L_B))$ has appeared. This
factor disappears quickly as the magnetic length decreases below the ALP
Compton wavelength $1/m_\phi$ -- we can interpret it as ALPs needing at least a
wavelength inside the magnetic field to convert into photons. 

This picture allows us to understand qualitatively another aspect of the
problem. One might be puzzled by the fact that a non-relativistic
ALP, even if it is at rest, can convert into a photon with a momentum $k=m_\phi$;
where does the momentum come from? The obvious explanation is that momentum is
not conserved at a sharp boundary condition, and in this example we have indeed
two, the mirror and the sharp transition between the magnetic field region and
the vacuum. In order to demonstrate this, let us consider an idealized but more
physical condition in which the magnetic field drops smoothly from a value
$|\mathbf B_0|$ to 0 in a length $\delta$ after $L_B$. The same
Eq.~\eqref{eq:outsideB} holds for a non-homogeneous $|\mathbf B(x)|$ which we will take as
\begin{equation}    
|\mathbf B(x<L)|=|\mathbf B_0| \quad ;\quad |\mathbf B(L<x<\delta)|=|\mathbf B_0|\(1-\frac{x-L}{\delta}\)
\quad ;\quad |\mathbf B(x>\delta)|=0 . 
\end{equation}    
The result is 
\begin{equation}
\label{eq:outsideBtwo}
\mathbf{E}(\mathbf{x},t)_{||}=
 \phi_0\,g \, {\mathbf B}_{0, ||}
\exp(-i m_\phi t)\exp(i m_\phi x) \(1-\frac{\sin\Delta}{\Delta}\cos(m_\phi L_B+\Delta)\), 
\end{equation}  
where $\Delta= m_\phi \delta/2$. If $\delta\to 0$ we recover the previous case
of a sharp boundary, but as $\delta$ grows large compared to $1/m_{\phi}$ the $\cos(m_\phi L_B+\Delta)$ factor gets suppressed. 
By considering the Fourier transform of a step function we realise that it has
indeed contributions from all momenta. However as we soften it the distribution
of momenta gets suppressed at length scales longer than $\delta$, the magnetic field boundary
is not able to produce momentum for the photons and drops out.  In
this limit we recover Eq.~\eqref{electricnear2}, except from the constant
electric field included in the ALP wave when it is inside of a magnetic field.
A smooth boundary is like having no boundary at all. 

At the practical level, we have seen that if the magnetic field transition is
smooth compared with the wavelength of interest, we can neglect the effects of
its boundary and Eq.~\eqref{electricnear2} holds (with the -1 if we are inside
the magnetic field or without it if we are outside). 

In the 3-D case the situation is more complicated but still we can get a
simple picture. Neglecting diffraction, the photons  emitted by the magnetic
field boundary are emitted perpendicularly to it. Then, if this is somewhat
irregular, as we expect in a realistic experiment, they will not interfere with
the ones emitted by the mirror (the other boundary that emits photons) and will
not be focused in the same place\footnote{In some frequency range it could be possible to use the
interference between the two to get a factor 2 enhancement in the amplitude (4
in power) at the price of designing a suitably magnetic field configuration.
However, for slightly different frequencies we shall have negative interference
instead. Since we have the goal of proposing broadband experiments we shall not
explore this possibility here.}.  

In the hidden photon case, the shielding boundary plays a similar role to the boundary of the magnetic field.
The two important points are: 1) The shielding typically will not focus the radiation generated by it in the same point (one can construct it that way). 2) Smoothly absorbing surfaces behave like a smooth magnetic field transition in that they generate little or no radiation.

\subsection{Effects of refraction and absorption}

It is well known that matter effects can affect tremendously the mixing of photons with low mass WISPs like hidden photons or ALPs. 
If the mixing angle is small, it gets modified as
\begin{equation}
\label{eq:effmixing}
\chi^2\to \chi_{\rm eff}^2=\chi^2\frac{m_{\gamma'}^4}{(m_{\gamma^\prime}^2+2\omega^2(n-1))^2+\mu^2}
\end{equation}
where $n=n(\omega)$ is the index of refraction and $\mu^2\approx {\rm max}\{\chi^2m_{\gamma'}^2,\omega \Gamma\}$. Here $\Gamma=\Gamma(\omega)$ is the photon absorption coefficient in the medium. 
Depending on the sign and magnitude of $n-1$ we can have enhanced or suppressed mixing%
\footnote{In a fully ionized plasma $n-1=-\omega_{\rm P}^2/2\omega^2$ and the enhancement can be quite large (we speak of resonant enhancement) for $\omega_P^2\sim m_{\gamma'}$ which can have very spectacular consequences~\cite{Jaeckel:2008fi,Arias:2012az}. 
Unfortunately, such a plasma is not the ideal medium to perform a low background experiment.}. 

Using $\omega=m_{\gamma'}$ we see that the significant suppression\footnote{Indeed an interesting way to understand the generation of radiation from the boundary is that radiation is generated when there is an abrupt change in $\chi_{\rm eff}$. This occurs at the mirror but also at a sharp magnetic field boundary where $\chi_{\rm eff}$ goes from 0 to a finite value. In general we expect radiation from boundaries where the complex index of refraction changes by an amount of order 1 in a wavelength $\sim 1/m_{\phi}$.}  of the mixing angle occurs only if $n-1$ is relatively large\footnote{Actually, the formula Eq.~\eqref{eq:effmixing} is not accurate for such large values $n-1$ but clearly conveys the qualitative features.} $n-1\gtrsim 0.5$ or if 
the factor $(m_{\gamma'}/\Gamma(m_{\gamma'}))^2\gg1$, i.e. when the absorption length is smaller than the Compton wavelength of the dark matter WISP.

In a realistic setup, we will typically have a residual non-ionized gas, like air, filling the space between the mirror and the detector. 
At normal conditions air has $n-1$ of order $10^{-4}$, and correspondingly small attenuation lengths and it will not reduce the sensitivity at all.

\section{Conclusions}{\label{sec:con}
In this letter we have described a new method for searching cold dark matter made out of axion-like particles or hidden photons.
The setup uses that reflective surfaces effectively convert these dark matter particles into electromagnetic radiation emitted perpendicular to the surface.
Using a spherical surface the emitted radiation is concentrated in the centre of the sphere where it can be detected.
The advantage over conventional resonant cavity searches is threefold. 
First the emitted power is proportional to the area of the surface and therefore easy to scale up. 
Second the setup is sensitive to a whole mass range in one measurement without the need to scan. 
Third it provides sensitivity also at higher masses which are difficult to access in cavity experiments.

\section{Acknowledgements}{\label{sec:ack}
  Javier Redondo gratefully acknowledges support by the European Union FP7 ITN
  INVISIBLES (Marie Curie Actions, PITN-GA-2011-289442), the Humboldt
  Foundation and the hospitality of Heidelberg University during the
  last stages of this project.  Andrei Lobanov acknowledges support from
  the Collaborative Research Center (Sonderforschungsbereich) SFB 676
  ``Particles, Strings, and the Early Universe'' funded by the German
  Research Society (Deutsche Froschungsgemeinschaft, DFG).


\begin{thebibliography}{99}

\bibitem{Jaeckel:2010ni}
  J.~Jaeckel and A.~Ringwald,
  Ann.\ Rev.\ Nucl.\ Part.\ Sci.\  {\bf 60} (2010) 405
  [arXiv:1002.0329 [hep-ph]].

\bibitem{Ringwald:2012hr}
  A.~Ringwald,
  arXiv:1210.5081 [hep-ph].

\bibitem{Arias:2012az}
  P.~Arias, D.~Cadamuro, M.~Goodsell, J.~Jaeckel, J.~Redondo and A.~Ringwald,
  JCAP {\bf 1206} (2012) 013
  [arXiv:1201.5902 [hep-ph]].
  
\bibitem{Nelson:2011sf}
  A.~E.~Nelson and J.~Scholtz,
  Phys.\ Rev.\ D {\bf 84} (2011) 103501
  [arXiv:1105.2812 [hep-ph]].

\bibitem{Zioutas:1998cc}
  K.~Zioutas, C.~E.~Aalseth, D.~Abriola, F.~T.~Avignone, III, R.~L.~Brodzinski, J.~I.~Collar, R.~Creswick and D.~E.~Di Gregorio {\it et al.},
  Nucl.\ Instrum.\ Meth.\ A {\bf 425} (1999) 480
  [astro-ph/9801176];
  K.~Zioutas {\it et al.}  [CAST Collaboration],
  Phys.\ Rev.\ Lett.\  {\bf 94} (2005) 121301
  [hep-ex/0411033];
%
  S.~Andriamonje {\it et al.}  [CAST Collaboration],
  JCAP {\bf 0704} (2007) 010
  [hep-ex/0702006];
%
  E.~Arik {\it et al.}  [CAST Collaboration],
  JCAP {\bf 0902} (2009) 008
  [arXiv:0810.4482 [hep-ex]];
%
  S.~Aune {\it et al.}  [CAST Collaboration],
  Phys.\ Rev.\ Lett.\  {\bf 107} (2011) 261302
  [arXiv:1106.3919 [hep-ex]].


\bibitem{Moriyama:1998kd}
  S.~Moriyama, M.~Minowa, T.~Namba, Y.~Inoue, Y.~Takasu and A.~Yamamoto,
  Phys.\ Lett.\ B {\bf 434} (1998) 147
  [hep-ex/9805026];
  %
  Y.~Inoue, T.~Namba, S.~Moriyama, M.~Minowa, Y.~Takasu, T.~Horiuchi and A.~Yamamoto,
  Phys.\ Lett.\ B {\bf 536} (2002) 18
  [astro-ph/0204388].

  
\bibitem{Robilliard:2007bq}
  C.~Robilliard, R.~Battesti, M.~Fouche, J.~Mauchain, A.~-M.~Sautivet, F.~Amiranoff and C.~Rizzo,
  Phys.\ Rev.\ Lett.\  {\bf 99} (2007) 190403
  [arXiv:0707.1296 [hep-ex]];
%
  A.~S.~.Chou {\it et al.}  [GammeV (T-969) Collaboration],
  Phys.\ Rev.\ Lett.\  {\bf 100} (2008) 080402
  [arXiv:0710.3783 [hep-ex]];
%
  K.~Ehret, M.~Frede, S.~Ghazaryan, M.~Hildebrandt, E.~-A.~Knabbe, D.~Kracht, A.~Lindner and J.~List {\it et al.},
  Phys.\ Lett.\ B {\bf 689} (2010) 149
  [arXiv:1004.1313 [hep-ex]];
%
  P.~Pugnat {\it et al.}  [OSQAR Collaboration],
  Phys.\ Rev.\ D {\bf 78} (2008) 092003
  [arXiv:0712.3362 [hep-ex]].

\bibitem{Jaeckel:2007ch}
  J.~Jaeckel and A.~Ringwald,
  Phys.\ Lett.\ B {\bf 659} (2008) 509
  [arXiv:0707.2063 [hep-ph]];
  A.~Wagner, G.~Rybka, M.~Hotz, L.~Rosenberg, S.~J.~Asztalos, G.~Carosi, C.~Hagmann and D.~Kinion {\it et al.},
  Phys.\ Rev.\ Lett.\  {\bf 105} (2010) 171801
  [arXiv:1007.3766 [hep-ex]];
%
  M.~Betz and F.~Caspers,
  Conf.\ Proc.\ C {\bf 1205201} (2012) 3320
  [arXiv:1207.3275 [physics.ins-det]];
%
  R.~Povey, J.~Hartnett and M.~Tobar,
  Phys.\ Rev.\ D {\bf 84} (2011) 055023
  [arXiv:1105.6169 [physics.ins-det]];
  R.~Povey, J.~Hartnett and M.~Tobar,
  Phys.\ Rev.\ D {\bf 82} (2010) 052003
  [arXiv:1003.0964 [hep-ex]];
  P.~H.~Williams, proceedings of the 6th Patras Workshop on Axions, WIMPs and WISPs 2010, http://axion-wimp2010.desy.de,
  DESY-PROC-2010-03.
  
\bibitem{Zavattini:2007ee}
  E.~Zavattini {\it et al.}  [PVLAS Collaboration],
  Phys.\ Rev.\ D {\bf 77} (2008) 032006
  [arXiv:0706.3419 [hep-ex]].
  
\bibitem{Sikivie:1983ip}
  P.~Sikivie,
  Phys.\ Rev.\ Lett.\  {\bf 51} (1983) 1415
   [Erratum-ibid.\  {\bf 52} (1984) 695].


\bibitem{Redondo:2008aa}
  J.~Redondo,
  JCAP {\bf 0807} (2008) 008
  [arXiv:0801.1527 [hep-ph]].


\bibitem{Schwarz:2011gu}
  M.~Schwarz, A.~Lindner, J.~Redondo, A.~Ringwald, G.~Wiedemann, 
  arXiv:1111.5797 [astro-ph.IM].

\bibitem{Irastorza:2011gs}
  I.~G.~Irastorza, F.~T.~Avignone, S.~Caspi, J.~M.~Carmona, T.~Dafni, M.~Davenport, A.~Dudarev and G.~Fanourakis {\it et al.},
  JCAP {\bf 1106} (2011) 013
  [arXiv:1103.5334 [hep-ex]].
  
\bibitem{Irastorza:2012qf}
  I.~G.~Irastorza {\it et al.}  [IAXO Collaboration],
  arXiv:1201.3849 [hep-ex].

\bibitem{Okun:1982xi}
  L.~B.~Okun,
  Sov.\ Phys.\ JETP {\bf 56} (1982) 502
   [Zh.\ Eksp.\ Teor.\ Fiz.\  {\bf 83} (1982) 892].

\bibitem{Redondo:2010dp}
  J.~Redondo and A.~Ringwald,
  Contemp.\ Phys.\  {\bf 52} (2011) 211
  [arXiv:1011.3741 [hep-ph]].


\bibitem{Asztalos:2009yp}
  S.~J.~Asztalos {\it et al.}  [ADMX Collaboration],
  Phys.\ Rev.\ Lett.\  {\bf 104} (2010) 041301
  [arXiv:0910.5914 [astro-ph.CO]].

\bibitem{Asztalos:2011ei}
  S.~J.~Asztalos, R.~Bradley, G.~Carosi, J.~Clarke, C.~Hagmann, J.~Hoskins, M.~Hotz and D.~Kinion {\it et al.},
  proceedings of the 7th Patras Workshop on Axions, WIMPs and WISPs 2011, based on a talk by G.~Rybka, http://axion-wimp2011.desy.de, DESY-PROC-2011-04.

\bibitem{Tada:1999tu}
  M.~Tada, Y.~Kishimoto, K.~Kominato, M.~Shibata, H.~Funahashi, K.~Yamamoto, A.~Masaike and S.~Matsuki,
  Nucl.\ Phys.\ Proc.\ Suppl.\  {\bf 72} (1999) 164.
  
\bibitem{Melissinos:2008vn}
  A.~C.~Melissinos,
  Phys.\ Rev.\ Lett.\  {\bf 102} (2009) 202001
  [arXiv:0807.1092 [hep-ph]].
  
\bibitem{Graham:2011qk}
  P.~W.~Graham and S.~Rajendran,
  Phys.\ Rev.\ D {\bf 84} (2011) 055013
  [arXiv:1101.2691 [hep-ph]].
  
\bibitem{Holdom:1985ag}
  B.~Holdom,
  Phys.\ Lett.\  B {\bf 166} (1986) 196; 
R.~Foot and X.~-G.~He,  
Phys.\ Lett.\ B {\bf 267} (1991) 509;   
  R.~Foot, H.~Lew and R.~R.~Volkas,
  Phys.\ Lett.\ B {\bf 272} (1991) 67.
  
\bibitem{Dienes:1996zr}
  K.~R.~Dienes, C.~F.~Kolda and J.~March-Russell,
  Nucl.\ Phys.\ B {\bf 492} (1997) 104
  [hep-ph/9610479];
  S.~A.~Abel, J.~Jaeckel, V.~V.~Khoze and A.~Ringwald,
  Phys.\ Lett.\ B {\bf 666} (2008) 66
  [hep-ph/0608248];
  S.~A.~Abel, M.~D.~Goodsell, J.~Jaeckel, V.~V.~Khoze and A.~Ringwald,
  JHEP {\bf 0807} (2008) 124
  [arXiv:0803.1449 [hep-ph]];
  M.~Goodsell, J.~Jaeckel, J.~Redondo and A.~Ringwald,
  JHEP {\bf 0911} (2009) 027
  [arXiv:0909.0515 [hep-ph]];
  M.~Cicoli, M.~Goodsell, J.~Jaeckel and A.~Ringwald,
  JHEP {\bf 1107} (2011) 114
  [arXiv:1103.3705 [hep-th]];
M.~Goodsell, S.~Ramos-Sanchez and A.~Ringwald, 
  JHEP {\bf 1201} (2012) 021
  [arXiv:1110.6901 [hep-th]].

\bibitem{Baker:2011na}
  O.~K.~Baker, M.~Betz, F.~Caspers, J.~Jaeckel, A.~Lindner, A.~Ringwald, Y.~Semertzidis, P.~Sikivie and K.~Zioutas,
  Phys.\ Rev.\ D {\bf 85} (2012) 035018
  [arXiv:1110.2180 [physics.ins-det]].

\bibitem{Irastorza:2012jq}
  I.~G.~Irastorza and J.~A.~Garcia,
  JCAP {\bf 1210} (2012) 022
  [arXiv:1207.6129 [physics.ins-det]].

\bibitem{Kogut:2012}
  A.~Kogut,
  ApJ {\bf 753} (2012) 110


\bibitem{Morrison:1977}
  P.~Morrison, J.~Billingham and J.~Wolfe
  NASA SP {\bf 419} (1977) 68


\bibitem{axiverse}
  J.~P.~Conlon,
  JHEP {\bf 0605} (2006) 078
  [hep-th/0602233];
  P.~Svrcek and E.~Witten,
  JHEP {\bf 0606} (2006) 051
  [hep-th/0605206];
  A.~Arvanitaki, S.~Dimopoulos, S.~Dubovsky, N.~Kaloper and J.~March-Russell,
  Phys.\ Rev.\ D {\bf 81} (2010) 123530
  [arXiv:0905.4720 [hep-th]];
  M.~Cicoli, M.~Goodsell, A.~Ringwald, M.~Goodsell and A.~Ringwald,
  JHEP {\bf 1210} (2012) 146
  [arXiv:1206.0819 [hep-th]].

\bibitem{Raffelt:1987im}
  G.~Raffelt and L.~Stodolsky,
  Phys.\ Rev.\ D {\bf 37} (1988) 1237.
  
\bibitem{Jaeckel:2008fi}
  J.~Jaeckel, J.~Redondo and A.~Ringwald,
  Phys.\ Rev.\ Lett.\  {\bf 101} (2008) 131801
  [arXiv:0804.4157 [astro-ph]].
%
  A.~Mirizzi, J.~Redondo and G.~Sigl,
  JCAP {\bf 0908} (2009) 001
  [arXiv:0905.4865 [hep-ph]].
  A.~Mirizzi, J.~Redondo and G.~Sigl,
  JCAP {\bf 0903} (2009) 026
  [arXiv:0901.0014 [hep-ph]].




\end{thebibliography}
\end{document}